# Experimental Gaussian Boson Sampling

Han-Sen Zhong[1,2], Li-Chao Peng[1,2], Yuan Li [1,2], Yi Hu[1,2], Wei Li[1,2], Jian Qin[1,2], Dian Wu[1,2], Weijun Zhang[3], Hao Li[3], Lu Zhang[3], Zhen Wang[3], Lixing You[3], Xiao Jiang[1,2], Li Li[1,2], Nai-Le Liu[1,2], Jonathan P. Dowling[2,4,5], Chao-Yang Lu[1,2,†], Jian-Wei Pan[1,2]

[1] *Hefei National Laboratory for Physical Sciences at Microscale and Department of Modern Physics, University of Science and Technology of China, Hefei, Anhui, 230026, China*
[2] *CAS Centre for Excellence and Synergetic Innovation Centre in Quantum Information and Quantum Physics, University of Science and Technology of China, Hefei, Anhui, 230026, China*
[3] *State Key Laboratory of Functional Materials for Informatics, Shanghai Institute of Microsystem and Information Technology (SIMIT), Chinese Academy of Sciences, 865 Changning Road, Shanghai 200050, China*
[4] *Hearne Institute for Theoretical Physics and Department of Physics and Astronomy, Louisiana State University, Baton Rouge, Louisiana 70803, USA*
[5] *NYU- ECNU Institute of Physics at NYU Shanghai, Shanghai 200062, China*
[†] *cylu@ustc.edu.cn*

**Abstract**

Gaussian Boson sampling (GBS) provides a highly efficient approach to make use of squeezed states from parametric down-conversion to solve a classically hard-to-solve sampling problem. The GBS protocol not only significantly enhances the photon generation probability, compared to standard boson sampling with single photon Fock states, but also links to potential applications such as dense subgraph problems and molecular vibronic spectra. Here, we report the first experimental demonstration of GBS using squeezed-state sources with simultaneously high photon indistinguishability and collection efficiency. We implement and validate 3-, 4- and 5-photon GBS with high sampling rates of 832 kHz, 163 kHz and 23 kHz, respectively, which is more than 4.4, 12.0, and 29.5 times faster than the previous experiments. Further, we observe a quantum speed-up on a NP-hard optimization problem when comparing with simulated thermal sampler and uniform sampler.

## Introduction:

Boson sampling is a promising route to demonstrate "quantum computational supremacy" over classical computers in the near term [1]. As a simplified non-universal quantum computing model, boson sampling only requires single photon sources, passive beam splitter networks and single-photon detectors [2]. The original boson sampler, proposed by Aaronson and Arkhipov, sends $n$ indistinguishable single photons into an $m$-mode ($m \sim n^2$) interferometer, and samples the photon number distribution of output states. The non-trivial regime for boson sampling is to be performed with a few tens of single photons to reach a high level of computational complexity [3].

Since 2011, there has been intense efforts towards demonstrating boson sampling [4–9], yet these all remain at small scales. The major experimental challenge of boson sampling is reliable sources of single photons, which is also at the heart of universal linear optical quantum computing [10] and the development of such sources has attracted on-going efforts [11–13]. Meanwhile, on the theoretical side, variant models of boson sampling with imperfect single-photon sources have been designed to increase the multiphoton count rate [14–18]. Lund *et al.* [14] proposed to use $\sim n^2$ probabilistic sources in the input modes, and sampled in the output $n$-photon events, yielding an exponential growth of possible input-output combination. In a similar spirit, Aaronson and Brod [15] put forward another variation—lossy boson sampling—to post-select $n$-photon events from input $n+k$ photons, where the sampling rate increases exponentially with $k$. However, it still remains an open question in the photon loss ($k$) boundary to retain the original complexity in standard boson sampling [19].

Recently, a new protocol called Gaussian boson sampling (GBS) [17,20,21] offers a unique shortcut to enormously enhance the generation efficiency based on parametric down-conversion (PDC). The GBS exploited the full Gaussian nature of the photon sources, while the previous works [4,5,9] relied on probabilistically post-selected single-photon Fock states from the Gaussian state naturally produced by PDC. By using the PDC as squeezed vacuum state sources rather than pseudo-single photon sources, we are no longer limited to low-gain regime, which favors implementations with the current technology. Moreover, GBS links to potential applications such as simulation of molecular vibrionic spectra [22], molecular docking [23] and graph theory [24–27].

The most crucial resource for performing the GBS is degenerate squeezed vacuum state with high efficiency and high purity, simultaneously. Possible frequency correlation in the PDC will either reduce the Hong-Ou-Mandel quantum interference visibility or reduce the efficiency by passive spectral filtering [28]. As in the lossy boson sampling case [15], the tolerable loss for the GBS to retain the #P-hard complexity is also an open question [17]. Nevertheless, it is clear that experimental efforts should be made to minimize the loss, for both retaining the fidelity of the Gaussian states and increasing the multi-photon rate.

**<u>Realization of GBS:</u>**

In our recent 12-photon entanglement experiment [9], we developed a wavelength-degenerate, frequency uncorrelated PDC source with near-unity collection efficiency, which immediately allows for a high-performance GBS experiment. The experimental setup is shown in Figure 1. The pumping Ti:Sapphire laser has a pulse duration of 160 fs and a central wavelength of 775 nm. By successively focusing the laser beam on three 6.3 mm-length $\beta$-barium borate (BBO) crystals with a pump beam waist of 260 $\mu$m, we obtain three pairs of two-mode squeezed vacuum sources without frequency correlation. We observe two-photon Hong-Ou-Mandel interference visibility of 0.99 with the o-ray and e-ray from the same BBO and 0.96 with the o-ray from independent BBOs.

The two spatial modes of the squeezed states, which have different polarizations, are coupled into a single-mode fiber and fed into an optical interferometer. Before and after the interferometer, half-wave plates and quarter-wave plates are inserted with random orientations to obtain a random transformation between different polarization modes. Together with six spatial modes and two polarization modes, this arrangement effectively produces a random 12×12 mode transformation. The measured amplitudes and phases of the transformation are plotted in Fig. 1b, c. Finally, we sample the

Gaussian state with 12 superconducting nanowire single-photon detectors with average detection efficiency of 75% at 1550 nm.

**Results of GBS:**

In contrast to standard Boson sampling, the relative input phases of each PDC can also affect the sampling distribution in the GBS. Firstly, we use only one pair of the two-mode squeezed state from the same PDC source with a fixed phase, a 0.99 quantum interference visibility and a squeezing parameter of 0.31, to verify our device works properly. We observe sampling rates of 690 kHz, 34 kHz and 3 kHz for the 2-, 3- and 4-photon GBS, respectively. To quantify the overlap between the experimental distribution ($p_i$) and theoretical distribution ($q_i$, see supplemental materials for theoretical analysis), we extract the similarity and the total variation distance, defined by $S = \sum_i \sqrt{q_i p_i}$ and $D = (1/2)\sum_i |p_i - q_i|$. The measured distances are 0.045(1), 0.064(1) and 0.081(2), and similarities are 0.998(1), 0.996(1) and 0.994(1) for the 2-, 3- and 4-photon GBS, respectively. The excellent agreement between theory and experiment implies the correct operation of the devices.

Next, we inject three pairs of two-mode squeezed state into the interferometer with squeezing coefficients of 0.365, 0.363 and 0.418, respectively. The measured sampling rates are 832 kHz, 163 kHz and 23 kHz for the 3-, 4- and 5-photon GBS, which is more than 4.4, 12.0, and 29.5 times faster than the previous experiments [4–9]. The experimental and theoretical distribution are plotted in Figure 2, from which we extract the similarities of 0.990(1), 0.982(1), 0.978(2) and the distances of 0.099(1), 0.147(2), 0.160(5) for 3-, 4- and 5-boson sampling, respectively.

To validate the GBS, we apply statistical methods to rule out several hypotheses, including the thermal state [29], distinguishable single photons [30], and a uniform sampler [30]. The average photon number of each input mode in thermal state hypothesis and distinguishable single photons hypothesis is equal to the experimental average photon number. Thermal state hypothesis can be understood as eliminating the entanglement of each two-mode squeezed state. Distinguishable single photons hypothesis is standard distinguishable sampling with photon loss and distinguishable single photons. All these hypotheses can be efficiently simulated by classical computer [29,30], and should be ruled out to verify that our experimental sampling is not easy to simulated. Firstly, we sort simulated thermal distributions, theoretical distributions and experimental results in Figure 3(a)–(c). The reordered experimental distributions are similar to the theoretical distributions but significantly steeper than the simulated thermal distributions, indicating non-classical interference in the interferometer. Secondly, we applied a modified likelihood ratio test [30] to exclude these hypotheses (see supplemental materials for details). As shown in Figure 3(d)–(f), we observe continuing deviation between our experimental Boson sampler and hypothesis sampler. The counter greater than zero means that the distribution of sampling is more similar to boson sampling than other hypothetical sampling. We reject all these hypotheses because of the increasing curves of experimental results.

**Discussion:**

Since GBS strongly relates to graph theory problems [24–27], we move to investigate the quantum speed-up for a NP-Hard optimization task named max-Haf problem [24]. The goal is to find a $k$-vertex subgraph with the largest Hafnian (in absolute value). Many optimization algorithms, such as simulated annealing and random search, require samples to update current result. In lossless photon-number-resolving GBS, the sampling probability is proportional to the Hafnian squared of correspond subgraphs [17], which result in an intrinsic advantage than classical sampling and can be used to improve optimization algorithms [24]. For any complex-weighted undirected graph, we can perform a GBS experiment where the top-left section of the sampling matrix is equal to the adjacent matrix of the graph [20]. We compare the performance of our experiment, a simulated thermal sampler, and a uniform sampler by using a random search algorithm, which is selecting subgraph from the given $n$ samples at random. These results are plotted in Figure 4. Here we normalize searching results by the exact solution. This diagram shows that GBS beats the simulated thermal sampler and uniform sampler. It is expected to observe more significant advantages are expected for a large-scale dense graph [24].

The high sampling rate is a key advantage of GBS [17]. To compare the computational speeds [1], we calculate the number of addition and multiplication operations on the classical computer for the same sampling task. We employ the sampling algorithm proposed by Quesada *et al.* with a complexity of $O(m^2 2^n)$ for sampling a *n*-click event from a *m*-mode gaussian state [21], which takes 7.4 k, 9.8 k and 13.0 k multiplication operations and 6.9 k, 9.0 k and 11.9 k addition operations in average to sample a 3-, 4- and 5-click event with 12 detectors, respectively. Thus, we estimate that a classical computer with the same sampling rate should calculate multiplications and additions at 8.1 GHz and 7.5 GHz respectively, which is comparable with the performance of the first generation of iPad. (see supplemental materials for details)

**Conclusion:**

In conclusion, we have demonstrated the first GBS experiment and its application in solving a NP-hard optimization problem. The sampling rate exceeds previous boson sampling experiment by about one order of magnitude. We estimate that the state-of-the-art PDC technology [9] can enable us to perform a large-scale GBS which can be faster than classical supercomputers. An open question is to investigate the quantum-to-classical transition in the GBS with photon loss.

**Acknowledgments:**

This work was supported by the National Natural Science Foundation of China (91836303, 11674308, and 11525419), the Chinese Academy of Sciences, the National Fundamental Research Program (2018YFA0306100) and the Anhui Initiative in Quantum Information Technologies.

## Appendix A. Supplementary materials

Supplementary materials to this article can be found online at https://doi.org/10.1016/j.scib.2019.04.007.

**Caption**

**Figure 1:** Experimental setup for GBS. (a) Three pairs of degenerate-frequency uncorrelated two-mode squeezed vacuum states are generated by three laser-pumped BBO crystals and coupled into a 12-mode optical interferometer, which includes six spatial modes and two polarization modes. From top to bottom the input modes are labelled as $\{1,4,2,5,3,6\}$, and the output modes are labelled from 1 to 12. Superconducting nanowire single-photon detectors are employed to perform threshold detection in each output mode. In (b) and (c) we show the amplitudes and phases of the measured transformation matrix.

**Figure 2**: Experimental (red) and theoretical (blue) distribution for 3-, 4- and 5-click GBS. The squeezing coefficients of two-mode squeezed vacuum states are 0.365, 0.363 and 0.418, respectively. The output combinations are denoted by $\{i,j,...\}$ where the detectors in output modes $i,j,...$ click. The similarities are measured to be 0.990(1),

0.982(1), 0.978(2) and the distances are measured to be 0.099(1), 0.147(2), 0.160(5) for 3-, 4- and 5-photon sampling, respectively.

**Figure 3**: Two methods are used to validate GBS. (a)–(c) Theoretical distribution (black), simulated thermal distribution (red) and experimental distribution (blue) in an ascending order for 3-, 4- and 5-click events, respectively. (d)–(f) Extended likelihood ratio tests reject hypotheses of thermal sampler (d), distinguishable single photon sampler (e) and uniform sampler (f). We assume that the input states of thermal sampler are thermal state with the same average photon number as experimental result, the distinguishable single photon sampler is standard distinguishable sampling with lossy distinguishable single phtotons and the uniform sampler is sampling with uniform distribution. The solid curves represent experimental results and the dashed curves represent simulated thermal/distinguishable single photon/uniform samples.

**Figure 4:** Quantum speedup on an approximate optimization max-Haf problem with random search algorithm. Searching results are obtained from experimental or simulated classical samples. We plot the average search results as a function of the number of samples. Points of experimental results are plotted with 490 k 4-photon samples for 4-subgraph and 8.4 k 6-photon samples for 6-subgraph. Each point of simulated thermal/uniform sampler is plotted with 500 k samples. Results are normalized by the maximum Hafnian of all subgraphs. The higher experimental curves indicate a quantum speed-up in this optimization problem.

# Figure 1

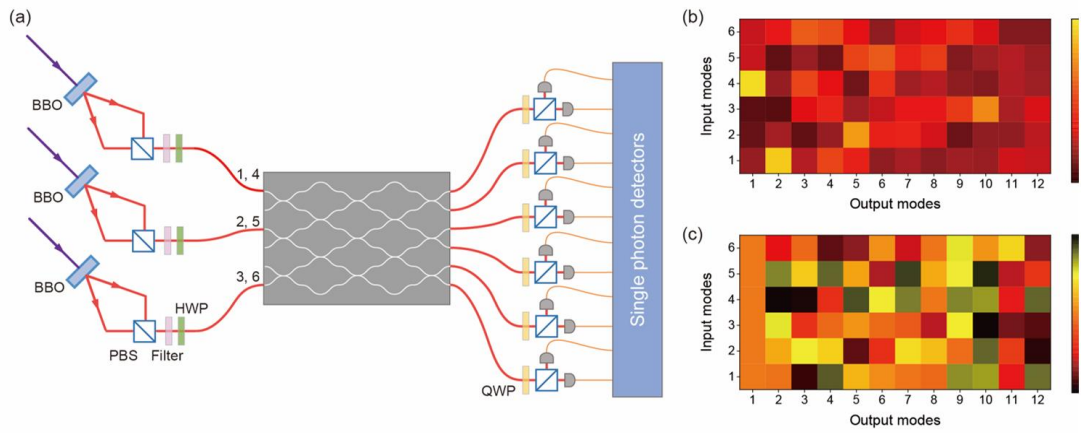

# Figure 2

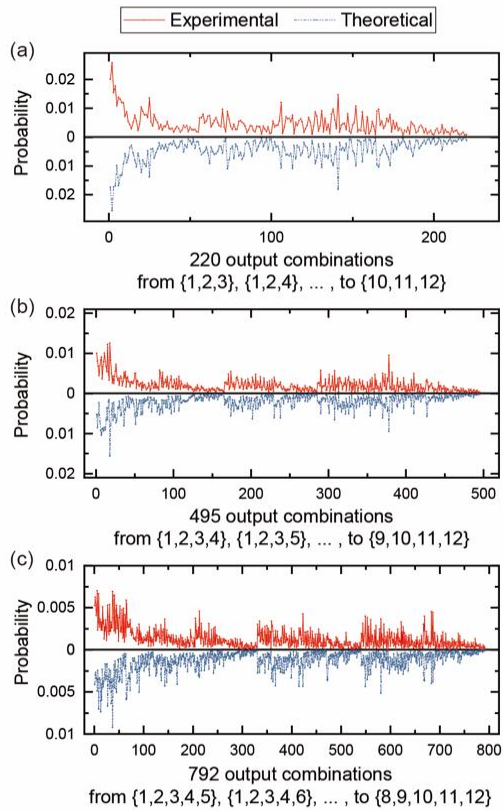

# Figure 3

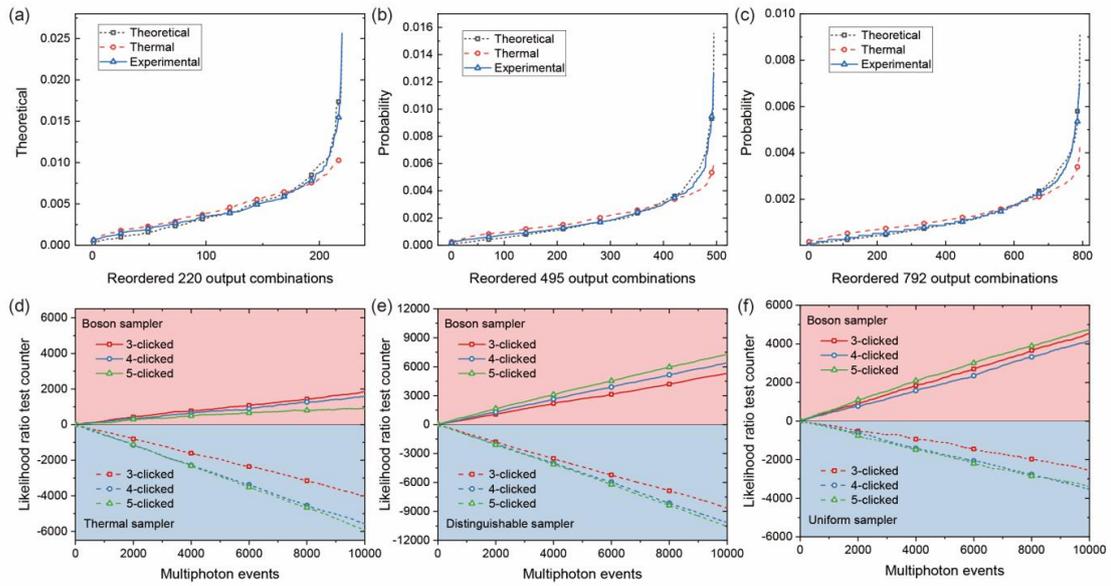

# Figure 4

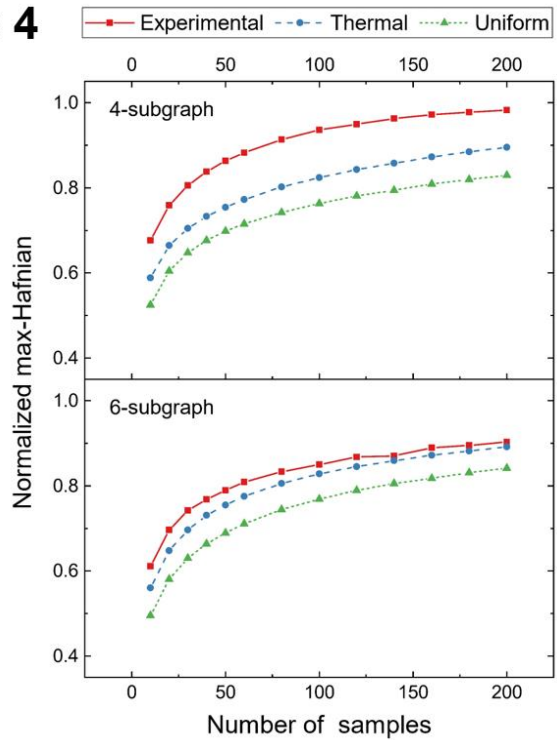